\documentclass[sigconf, nonacm, pdfa]{acmart}

\usepackage{graphicx}
\usepackage{balance}  
\usepackage{epsfig}
\usepackage{graphicx}
\usepackage{epstopdf}
\usepackage{latexsym}
\PassOptionsToPackage{noend}{algpseudocode}
\usepackage{algpseudocode}
\usepackage{pifont}
\usepackage{epsfig}
\usepackage{amssymb}
\usepackage{amsmath}
\usepackage{amsfonts}
\usepackage{subfigure}
\usepackage{stmaryrd}
\usepackage{url}
\usepackage{multirow}
\usepackage{array}
\usepackage[normalem]{ulem}
\usepackage{color}
\usepackage{cleveref}
\usepackage{xspace}
\usepackage{mathtools}
\usepackage{soul}
\usepackage{listings}
\usepackage{enumitem}
\usepackage{xcolor}
\usepackage{tikz}
\newsavebox{\blackball}
\newsavebox{\greenball}

\usepackage[utf8]{inputenc}
\usepackage{enumitem}
\usepackage{amsmath}

\usepackage{bbding}
\usepackage{amssymb}
\usepackage{epstopdf}
\usepackage{epsfig,endnotes}
\usepackage{url}
\usepackage{array}
\usepackage{booktabs}
\usepackage{threeparttable}
\usepackage{tabulary}

\usepackage{CJKutf8}
\usepackage[most,breakable]{tcolorbox}
\usepackage{etoolbox}
\usepackage{fancyhdr}
\usepackage{lipsum}
\usepackage[fencedCode]{markdown}

\usepackage[lined,boxed,vlined,ruled,linesnumbered]{algorithm2e}

\usepackage{bm}

\theoremstyle{definition}

\newcommand{\zxh}[1]{\textcolor{orange}{ #1}}

\newcolumntype{M}[1]{>{\centering\arraybackslash}m{#1}}
\errorcontextlines\maxdimen

\newcommand{\oursys}{\texttt{LLMDB}\xspace}

\newcommand{\llm}{\textsc{LLM}\xspace}
\newcommand{\llms}{\textsc{LLMs}\xspace}

\newcommand{\lvc}{\textsc{LVC}\xspace}

\newcommand{\vdb}{\textsc{VDB}\xspace}

\newcommand{\languagemodels}{{large language models}\xspace}

\newcommand{\hi}[1]{\vspace{.25em} \noindent {\bf #1}\xspace}

 \newcommand{\bfit}[1]{\textbf{\textit{#1}}}

\definecolor{codegreen}{rgb}{0,0.6,0}
\definecolor{codegray}{rgb}{0.5,0.5,0.5}
\definecolor{codepurple}{rgb}{0.58,0,0.82}
\definecolor{backcolour}{rgb}{0.95,0.95,0.92}

\lstdefinestyle{mystyle}{
	backgroundcolor=\color{backcolour},   
	commentstyle=\color{codegreen},
	keywordstyle=\color{magenta},
	numberstyle=\tiny\color{codegray},
	stringstyle=\color{codepurple},
	basicstyle=\ttfamily\footnotesize,
	breakatwhitespace=false,         
	breaklines=true,                 
	captionpos=b,                    
	keepspaces=true,                 
	numbers=left,                    
	numbersep=5pt,                  
	showspaces=false,                
	showstringspaces=false,
	showtabs=false,                  
	tabsize=2
}
\lstdefinestyle{jsonStyle}{
	basicstyle=\small\ttfamily,
	columns=fullflexible,
	showstringspaces=false,
	commentstyle=\color{codegreen}\upshape,
	stringstyle=\color{codegreen},
	morestring=[b]",
	moredelim=[s][\color{codepurple}]{\{}{\}},
	moredelim=[s][\color{codepurple}]{[}{]},
	moredelim=[l][\color{codepurple}]{:},
	moredelim=[l][\color{codepurple}]{,}
}

\begin{document}

\title{LLM As Database Administrator}
\title{\oursys: A System that Unifies Data and Model Management}
\title{\llm-Enhanced Data Management [Vision]}

\author{Xuanhe Zhou}
\affiliation{%
	\institution{Tsinghua University}
	\city{}
	\country{}
}
\email{zhouxuan19@mails.tsinghua.edu.cn}

\author{Xinyang Zhao}
\affiliation{%
	\institution{Tsinghua University}
	\city{}
	\country{}
}
\email{xy-zhao20@mails.tsinghua.edu.cn}

\author{Guoliang Li}
\affiliation{%
	\institution{Tsinghua University}
	\city{}
	\country{}
}
\email{liguoliang@tsinghua.edu.cn}

\pagestyle{plain}
\pagenumbering{arabic}


\begin{abstract}
\begin{sloppypar}
Machine learning (ML) techniques for optimizing data management problems have been extensively studied and widely deployed in recent five years. However traditional ML methods have limitations on generalizability (adapting to different scenarios) and inference ability (understanding the context).  Fortunately, large language models (\llms) have shown high generalizability and human-competitive abilities in understanding context, which are promising for data management tasks (e.g., database diagnosis and data analytics). However, existing \llms have several limitations: hallucination,  high cost, and low accuracy for complicated tasks. To address these challenges, we design \oursys, an \llm-enhanced data management paradigm which has good generalizability and high inference ability while avoiding hallucination, reducing \llm cost, and achieving high accuracy. \oursys embeds domain-specific knowledge to avoid hallucination by \llm fine-tuning and prompt engineering. \oursys reduces the high cost of \llms by vector databases which provide semantic search and caching abilities. \oursys improves the task accuracy by \llm agent which provides multiple-round inference and pipeline executions. We showcase three real-world data management scenarios that \oursys can well support, including query rewrite, database diagnosis and data analytics. We also summarize the open research challenges of \oursys.
\end{sloppypar}
\end{abstract}

\maketitle


\vspace{-.5em}

\section{Introduction}
\label{sec:intro}


Machine learning algorithms have been widely adopted to optimize data management problems, e.g., data cleaning~\cite{tang2020rpt}, data analytics~\cite{DBLP:journals/tkde/ChaiLFL21}, query rewrite~\cite{DBLP:journals/pvldb/ZhouLCF21}, database diagnosis~\cite{DBLP:journals/pvldb/MaYZWZJHLLQLCP20}. However, traditional machine learning algorithms cannot address the generalizability and inference problem. For example, existing machine algorithms are hard to adapt to different databases, different query workloads, and different hardware environments. Moreover, existing machine algorithms cannot understand the context and do multi-step reasoning, e.g., database diagnosis for problem detection and root-cause analysis. Fortunately, large language models (LLMs) can address these limitations and offer great opportunities for data management~\cite{prompt,awesome-chatgpt-prompts,enhance-chatgpt}. For example, LLMs can be used to diagnose the database problems and help DBAs to find the root causes of slow SQL queries. \llms can also enable natural language (NL) based data analytics such that users can use NL to analyze their datasets. 


However \llms have three limitations: hallucination, high cost, and low accuracy for complicated tasks. These shortcomings pose significant challenges, even risks, especially when \llms are used in critical data management steps (e.g., price trend analysis in financial forecasting). Although there are some \llm agent based methods to overcome these challenges, like chains of thought~\cite{wei2022chain,cot,react,reflexion} and tool-calling functionalities~\cite{toolformer}.  These works, while impressive, reveal several limitations. First, they heavily rely on \llms to support nearly every tasks (e.g., designing, coding, and testing in \llm-based software development~\cite{qian2023communicative,hong2023metagpt}), leading to instability and a high error rate. For instance, GPT-4 might fail to extract text from slides due to misunderstandings of intent, despite having this capability. Second, for complex tasks like tool calling, extensive training data for specific APIs are required to fine-tune the \llm. This approach is vulnerable to API changes and can result in significant cost inefficiencies~\cite{qin2023toolllm}. Third, \llm agents still lack the capability to fully utilize knowledge from multiple sources, which are vital to mitigate hallucination issues in \llms (e.g., serving as evaluation criteria). Thus it calls for a new \llm-enhanced data management paradigm to address these limitations. 



\begin{figure}[!t]
	\vspace{.5em}
	\centering
	\includegraphics[width=.98\linewidth, trim={0 0 0 0},clip]{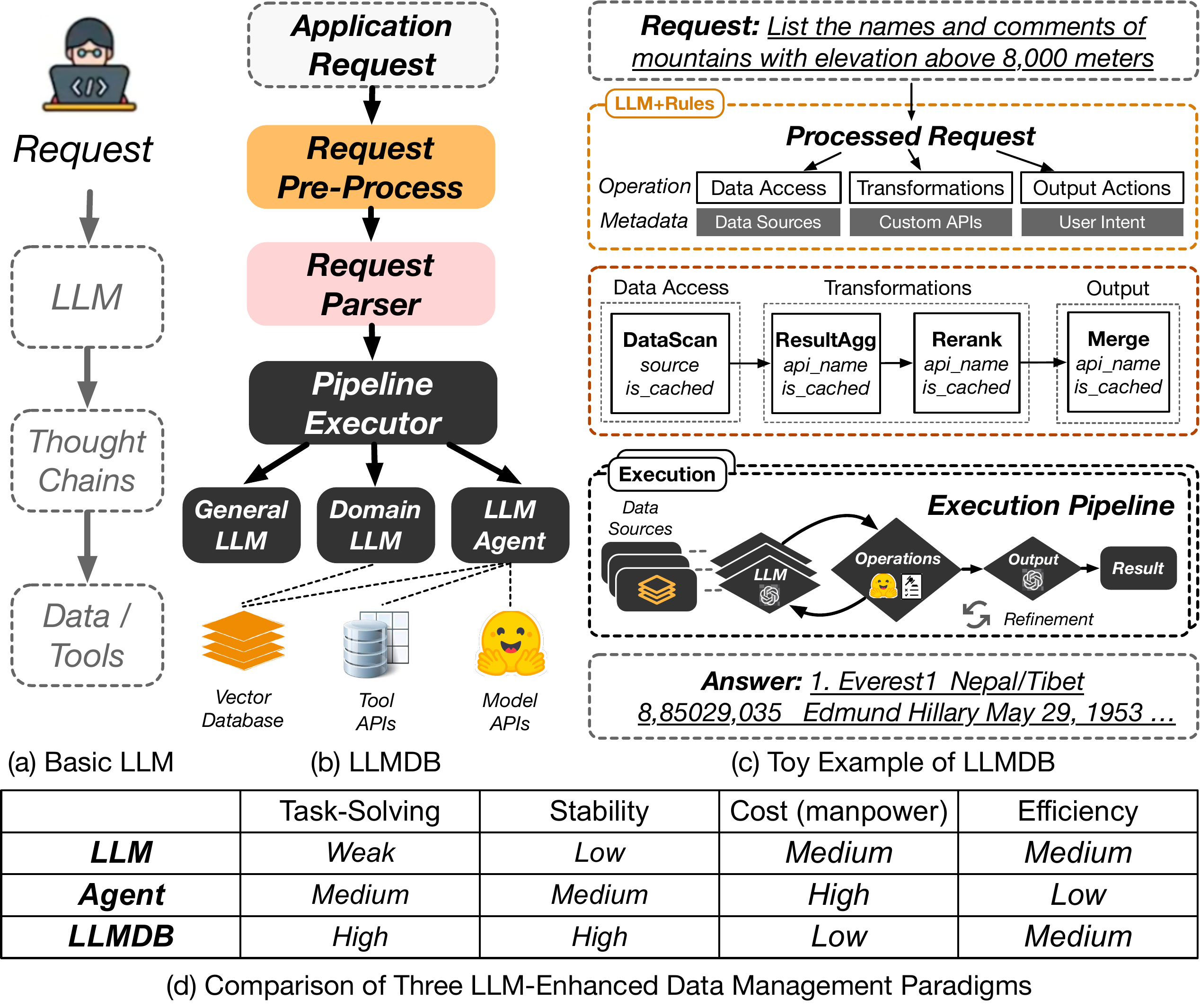}
	\vspace{-1.em}
	\caption{\llm for Data Management}
	\label{fig:intro}
	\vspace{-2.5em}
\end{figure}

\noindent \bfit{Vision of \llm-Enhanced Data Management.} Developing a robust  \llm-enhanced data management system is essential to harness the full potential of \llms. There are three main challenges.  First, {\it how to effectively utilize data sources} (e.g., tabular data and various document formats) to reduce \llm hallucination problems (e.g., through knowledge-augmented answering). Second, {\it how to reduce the \llm overhead?} It is rather expensive to call \llms for every request. It is important to accurately interpret the intent behind user requests and capture the domain knowledge in order to reduce the iterations with \llms. Third, the execution of complex data management tasks may involve multiple operations. \textit{How to efficiently manage these operations and pipelines} to enhance both execution effectiveness and efficiency.


\noindent \bfit{Idea of \oursys.} To address these challenges, we propose \oursys, a general framework for designing and implementing \llm-enhanced data management applications. As shown in Figure~\ref{fig:intro}, \oursys improves the basic \llm by incorporating a series of modular components. \oursys embeds domain-specific knowledge to avoid hallucination by \llm fine-tuning and prompt engineering. \oursys reduces the high cost of \llms by vector databases which provide semantic search and caching abilities. \oursys improves the task accuracy by \llm agent which provides multiple-round inference and pipeline executions.  
Specifically, \oursys leverages a combination of {\it general \llms, domain-specific \llms, \llm agent, vector databases, and data sources} to handle data management tasks. \llms provide understanding and inference ability. Users provide domain-specific knowledge and data sources. Domain-specific \llms fine-tune the general \llms based on the user-provided data sources, and provide the domain-specific knowledge. Vector databases generate embeddings of user-provided data sources, retrieve top-$k$ document segments, take them as prompts and input them to \llms. With the domain-knowledge prompts,  \llms can provide better answers. Moreover, for complicated tasks, \llm agent generates a pipeline with multiple operations to process the task.


\begin{figure*}[!t]
	\vspace{-.5em}
	\centering
	\includegraphics[width=1\linewidth, trim={0 0 0 0},clip]{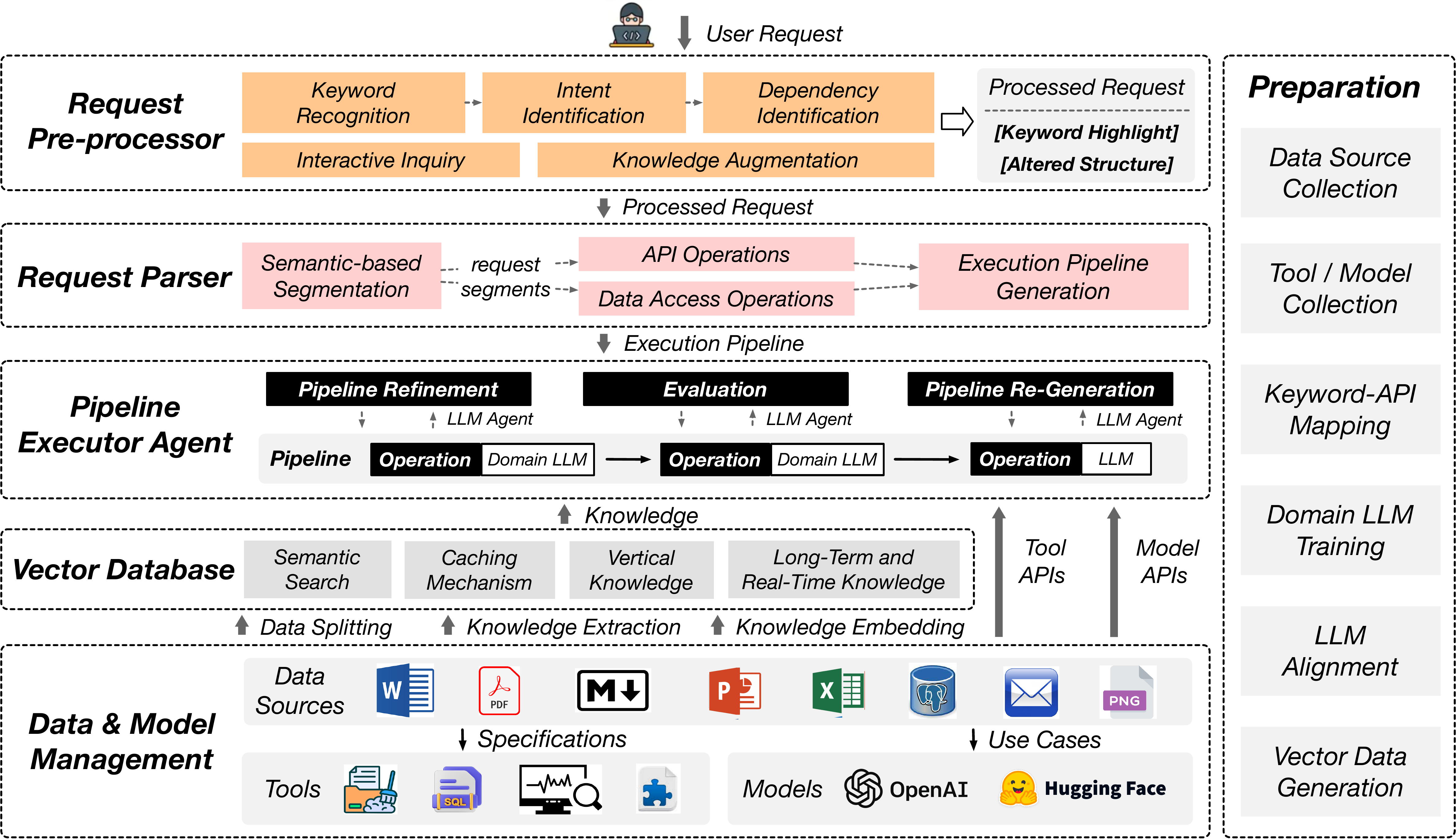}
	\vspace{-1.75em}
	\caption{\oursys Overview}
	\label{fig:arch}
	\vspace{-1em}
\end{figure*}

\hi{Contributions.} We make the following contributions.

\noindent (1) We design an \llm-enhanced data management system, which has generalizability and high inference ability while avoiding hallucination, reducing \llm cost, and achieving high accuracy.

\noindent  (2) We propose a domain-knowledge embedding and semantic search method to address the hallucination problem. We adopt vector databases to reduce the \llm overhead. We design a pipeline generation and execution framework to optimize complicated tasks. 



\noindent (3) We conduct case study on three typical data management applications using \oursys, and discuss the system advantages.

\noindent (4) We summarize research challenges of \llm-enhanced data management, including \oursys operation design, pipeline executor optimization, embedding selection, and knowledge augmentation.

\vspace{-1em}
\section{\oursys Paradigm}
\label{sec:overview}
\vspace{-.25em}

\subsection{\oursys Architecture} 

\oursys has five important components, including {\it general \llms, domain-specific \llms, \llm executor agent, vector databases, and data source manager}. Specifically, general \llms provide understanding and inference ability.  Data source manager provides domain-specific data and knowledge. Domain-specific \llms fine-tune the general \llms based on the user-provided data sources, which are used for request pre-processor and parser that capture the user intent and generate execution pipeline. Vector databases generate embeddings of user-provided data sources, retrieve top-$k$ document segments for user requests, take them as prompts and input them to \llms. With the domain-knowledge prompts,  \llms can provide high-quality results. Moreover, for complex tasks, \llm agent generates a pipeline with multiple operations to optimize the complex task. \oursys includes offline preparation and online inference.

\hi{Offline Preparation.} Given an application requirement, \oursys first collects and analyzes the information from various data sources. Then \oursys derives detailed specifications that outline the application objectives, traditional tools (e.g., various database plugins), and the usage scenarios of domain AI models (e.g., Marcoroni-7B model for sentiment analysis~\cite{Marcoroni}).

\hi{Online Inference.} For a user request, \oursys first parses the request to detect the intent and operations. Following this, a pipeline of the operations is generated, which is then executed in a graph-style format to ensure efficient processing. The final step involves an iterative process of result evaluation and pipeline re-generation, ensuring the accuracy and relevance of final outcome.


\subsection{Offline Preparation} 

Offline preparation is to initialize \llm-based data management applications before online usage. This involves five main modules.

{\it (1) Data Source Collection:} This module gathers a variety of data sources. For instance, in diagnosis task, the data can come from maintenance documents, historical logs, and operational records.

{\it (2) Tool/Model Collection:} Based on the collected data, this module gathers and sets up essential tools and small-scaled AI models. For instance, in database diagnosis, this might include the collection of monitoring tools, analytics tools, and image-to-text (e.g., for program running screenshot) models like vit-gpt2~\cite{vitgpt}. 


{\it (3) Keyword-API Mapping:} This module creates a mapping table (metadata) that links keywords or phrases to relevant tool or model APIs. For example, the keyword ``image classification'' might be mapped to an API of a convolutional neural network (CNN) model. This mapping enables the automated selection of appropriate tools or models based on the specific requirements of the input request.

{\it (4) Domain \llm Training:} This module fine-tunes small-scale domain models using the updated data sources, which ensures that the models are well-adapted to the application’s needs, leading to improved accuracy and performance in the target scenarios.

{\it (5) \llm Alignment:} This module adapts \llm workers to stuffs like interface language updates. For instance, an \llm worker may need to transition from SQL to a knowledge graph language, emphasizing semantic relationships over structured queries for richer, more complex data interactions.

{\it (6) Vector Data Generation:} This module generates embedding for each data source and inserts the embeddings into vector databases. The vector database provides the vector search ability.






\subsection{Online Inference}

Given the prepared {\it tool and model APIs/data sources/metadata}, next we explain how to handle various online requests.

\hi{Request Pre-processor.} This module prepares the input request so as to ease the following parsing and execution procedures. 

{\it (1) Keyword Recognition:} With the keyword corpus, it emphasizes keywords occurring within the input request, where we can use techniques like synonym matching (e.g., ``examination'' $\rightarrow$ ``analysis'') and stem matching (e.g., ``computing'' $\rightarrow$ ``comput-.'')~\cite{DBLP:conf/coling/Turney08}.

{\it (2) Intent Identification:} Apart from basic keyword recognition, it identifies the underlying purpose of a request using semantic analysis. For instance, consider the request \textit{``Identify similar property listings and compare their features''}. A semantic model will not only focus on keywords like ``compare'' and ``features'' but also understand the contextual meaning of ``similar property listings''. This involves identifying the implicit criteria for similarity (e.g., location, price range) and relevant features for comparison (e.g., number of bedrooms). Thus, the request is transformed into a more actionable form, such as \textit{``Retrieve property listings based on [criteria for similarity] from $\cdots$. Compare their [feature 1], [feature 2], $\cdots$''}. 

{\it (3) Dependency Identification} analyzes the grammatical structure to understand how keywords relate to each other, which is vital to the following {\it semantic-based request segmentation}.

{\it (4) Interactive Inquiry} allows the application to ask follow-up questions so as to obtain more information that may clarify the user's intent and required operations (e.g., more entity attributes).

{\it (5) Knowledge Augmentation} enriches the request with valuable knowledge (e.g., the meaning of $Prometheus$ in diagnosis task) or empirical rules (e.g., the criteria of entity matching).	


\hi{Request Parser.} This module interprets the user's request into an executable operation pipeline. 

{\it (1) Semantic-based Request Segmentation:} Different from basic techniques like \textit{Tokenization}, it splits the input request into meaningful segments. For example, with the Semantic Role Labeling (SRL) technique~\cite{marquez2008semantic}, it assigns roles to words in the pre-processed request to capture their semantic relationships. For instance, in the request {\it ``Find flights from New York to London departing tomorrow''}, it identifies segments such as {\it ``Find flights''}, {\it ``from New York to London''}, and {\it ``departing tomorrow''}, each conveying a specific operation or parameters for the flight search request.

{\it (2) Operation Mapping:} Based on the keyword-API mapping table, we convert the segments into functional operations (using tool or model APIs) and data access operations (using data sources). 

{\it (3) Execution Pipeline Generation:} These operations form a basic execution pipeline based on their structural logic of the request. For example, for a segmented request $R$ = [{\it ``Identify missing value''}, {\it ``in dataset column''}], $R_0$ is designated as the root, helping arrange other operations based on both the order and semantic relations.


\hi{Executor Agent.} Given the basic pipeline, this module refines the pipeline with underlying information, evaluates the outcomes, and re-generates more effective pipeline (arriving time or cost limit).

{\it (1) Pipeline Refinement:} For each operation in the pipeline, it enriches with information like $(i)$ which data source to use (e.g., the same document can be stored in both vector database and elastic search engine~\cite{gormley2015elasticsearch}) and $(ii)$ which operation orders are most execution-efficient (using methods like genetic algorithm). 

{\it (2) Evaluation:} It assesses the quality of results by executing the current pipeline. For instance, in a data analytics task, we can utilize an \llm to analyze user feedback about the usability and relevance of the analytics results.

{\it (3) Pipeline Re-Generation:} If the evaluation result is poor, it generates a new execution pipeline. For example, we can use decision trees and neural networks to identify and add any intermediate operations that are potential to derive more reliable results.

\hi{Vector Databases.}  Vector databases are used to improve the execution effectiveness and efficiency from multiple aspects.

{\it (1) Semantic Argumentation and Search:} When a knowledge query is issued by \llm worker, the vector database $(i)$ integrates context and intent analysis to enrich the query; and $(ii)$ utilizes advanced similarity search algorithms (e.g., graph-network-based embedding for relational knowledge patterns) to ensure high relevance in search results by understanding the semantic difference of queries.

{\it (2) Caching Mechanism:}  We can cache hot user requests and their answers, and when a similar request is posted, we can use vector database to directly answer the request, without involving \llms.


{\it (3) Vertical Knowledge:} It serves as a metadata container, enabling unified access and management of multiple data sources. 

{\it (4) Long-term and Real-time Knowledge:} It transforms time-series data into vector embeddings, facilitating the storage and retrieval of real-time and long-term data. This way, it enhances the system's ability to perform complex analysis over extended periods.

\hi{Data \& Model Management.} This module functions as the intermediary between \emph{Pipeline Executor} and data sources, tools, and AI models. It includes \llm agents (used during pipeline scheduling and execution), which serve as interfaces to external \textit{tool or model APIs}. This module $(i)$ aligns the system's internal understanding with external knowledge and $(ii)$ mitigates over-reliance on \llms.

\subsection{Challenges}
\label{subsec:cha:sec2}

There are several research challenges in this \oursys paradigm. First, how to effectively understand the user request and generate the execution pipeline? Second, how to select well-designed execution operations with completeness and minimalism such that they can be combined to generate high-quality execution pipeline? Third, how to design high-quality executor agent that can utilize multiple operations to effectively answer a complicated task? Fourth, how to select effective embedding method that can capture the domain-specific proximity such that can enrich the query with semantic augmentation? Fifth, how to balance \llm fine-tuning and prompt engineering?  Fine-tuning can optimize \llm with domain-specific knowledge, but it requires a large volume of high-quality training data; while prompt does no require training data, but it requires to add the prompts for every request. Sixth, how to utilize the \oursys feedback to iteratively optimize our pipeline.

\begin{figure}[!t]
	\vspace{.5em}
	\centering
	\includegraphics[width=.95\linewidth, trim={0 0 0 0},clip]{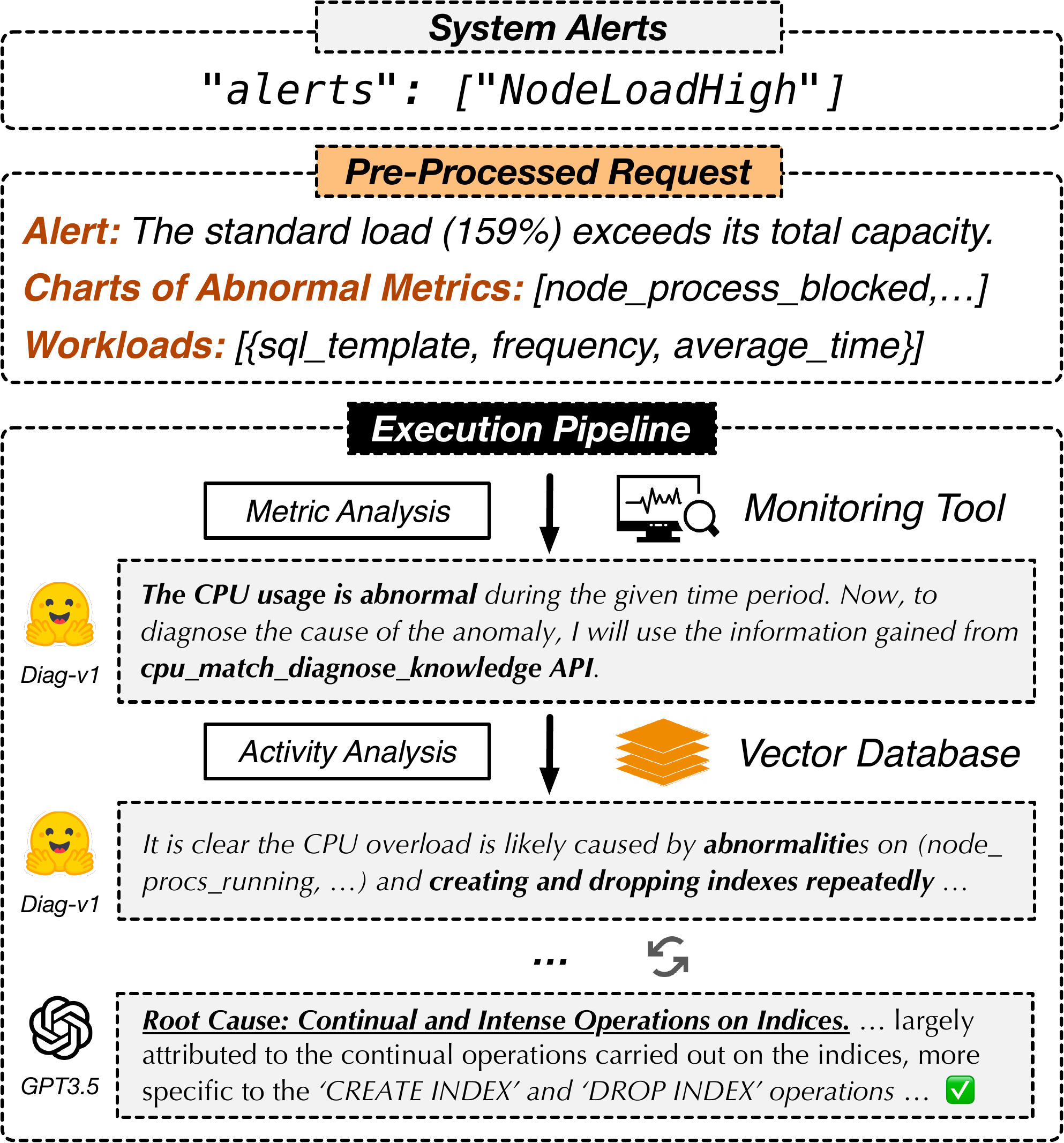}
	\vspace{-1em}
	\caption{\oursys for Database System Diagnosis}
	\label{fig:diagnosis}
	\vspace{-2em}
\end{figure}


\section{\oursys for System Diagnosis}
\label{sec:diagnosis}

Database system diagnosis focuses on identifying the root causes of anomalies, which inputs various monitoring information and outputs root causes and solutions to fix the root causes. Traditional machine learning methods often treat system diagnosis as a classification problem~\cite{DBLP:journals/pvldb/MaYZWZJHLLQLCP20,DBLP:conf/sigmod/YoonNM16,DBLP:conf/cidr/DiasRSVW05}. These methods $(i)$ rely heavily on large sets of high-quality labeled data, limiting their generalizability to new scenarios (e.g., with different monitoring metrics) and $(ii)$ cannot generate detailed analysis reports like human DBAs. While there are prototypes leveraging \llms to solve this problem~\cite{zhou2023llm,DBLP:journals/corr/abs-2312-01454}, their implementation requires substantial expertise. In contrast, we explore how to automate the development of \llm-enhanced database system diagnosis.

\hi{Domain-Specific Data Preparation.} First, \oursys extracts knowledge from extensive documents, including technical manuals, logs, and case studies. This extracted knowledge is then systematically structured and stored in vector database, facilitating efficient retrieval and analysis. Second, \oursys sets up a suite of diagnosis tools, each with $(i)$ defined APIs, $(ii)$ text specifications, and $(iii)$ demonstration examples, which are tailored scenarios illustrating the application of these tools in real-world system diagnosis. They not only serve as a guide for understanding the practical application of the tools, but also assist in fine-tuning \llms and domain models.

\hi{User Request Pre-Processing.} \oursys enriches original system alerts to create a comprehensive context for the following analysis. For instance, if an alert indicates high CPU usage, \oursys further decides relevant factors like {\it the duration of high usage}, {\it related processes consuming excessive CPU resources}, and {\it any concurrent system activities} that might contribute to the anomaly. Additionally, this stage includes workload or slow queries from the system (e.g., logs or system views) during the anomaly period. 


\hi{Request Parsing.} Since the execution pipeline of database system diagnosis is relatively fixed, \oursys directly maps the enriched alert into a pipeline with operations in the order of $(i)$ high-level metric analysis, $(ii)$ finer-grained metric analysis, $(iii)$ activity analysis, $(iv)$ workload analysis, and $(v)$ solution recommendation. Each operation is accordingly matched with relevant tool APIs (e.g., {\it index advisor API} for solution recommendation) and knowledge.

\hi{Execution Pipeline.} The execution pipeline for database diagnosis is primarily focused on two areas: tool callings for gaining specific abnormal status and knowledge-matching-based root cause analysis. First, tool calling sequences perform automated checks on database performance and integrity. For example, one operation may use a performance monitoring tool to identify slow queries, followed by an integrity checker for detecting data corruption. Second, knowledge matching in one operation leverages a vast repository of documented issues and solutions, comparing current anomalies with historical incidents to identify matches or similarities, which is addressed by vector databases. Moreover, there are also operations that need to implement selected solutions. For instance, a pattern of locking conflicts could lead to actions like $(i)$ adjusting transaction isolation levels or $(ii)$ redesigning the database schema.


\hi{Model Management.} Database diagnosis seldomly has an open corpus. Thus, it is vital to train \llm to learn anomaly root causes from real cases. This involves unsupervised \llm training to recognize and interpret complex patterns from sampled time-series data. For example, an \llm needs to identify that recurring divergences in transactional data could suggest concurrency conflicts. Similarly, \llm needs to analyze log files showing intermittent database access failures and correlate this with issues like network instability.


\hi{Research Challenges.} First, how to involve human experts in the diagnosis loop. Human experts may provide a vast number of valuable advice when using the diagnosis application in real cases, which should be learned by caching in vector database or fine-tuning the models. Second, how to integrate multi-modal information sources like text, figures, and videos in diagnosis process. This could provide a more comprehensive context and improve the accuracy of root cause identification.

\section{\oursys for Data Analytics}
\label{sec:data-analytics}

In many data analytical scenarios, users are unable to write SQL queries and prefer to use natural language (NL) to analyze the data.  A challenge here is how to generate analytical results from NL queries. Current \llm works~\cite{qin2023tool,chen2023seed} depend on \llms for crafting these queries, which often leads to errors (e.g., non-executable queries or results that do not meet user requirements). Thus, we showcase how to accurately conduct data analysis with \oursys.

\hi{Domain-Specific Data Preparation.} For a data analytic task like {\it ``You are given a table on $\cdots$. Filter the table to show only facilities located in Boston.''}, it mainly aims to generate executable analysis programs with visualizations as outputs. Thus, we prepare $(i)$ data sources like existing analysis requests and example programs in different languages (e.g., SQL, Python, R); and $(ii)$ models that are good at translating corresponding languages. For the data source preparation, we can adopt active learning to refine the \llm's understanding, focusing on generating diverse examples that cover a wide array of analytics scenarios. For model preparation, we can fine-tune \llms with training examples that involve Natural Language (NL) instructions, sequences of API calls (code lines), and detailed explanations.

\hi{Request Pre-Processing.}  When handling an incoming analytics request, we first prepare and load the raw tabular data. For example, when processing a dataset with attributes such as sales records, we standardize date formats by writing a program, typically in Python, that utilizes a series of Pandas functions like \emph{drop\_duplicates()} and \emph{pd.to\_datetime()}.

\begin{figure}[!t]
	\vspace{-.5em}
	\centering
	\includegraphics[width=.95\linewidth, trim={0 0 0 0},clip]{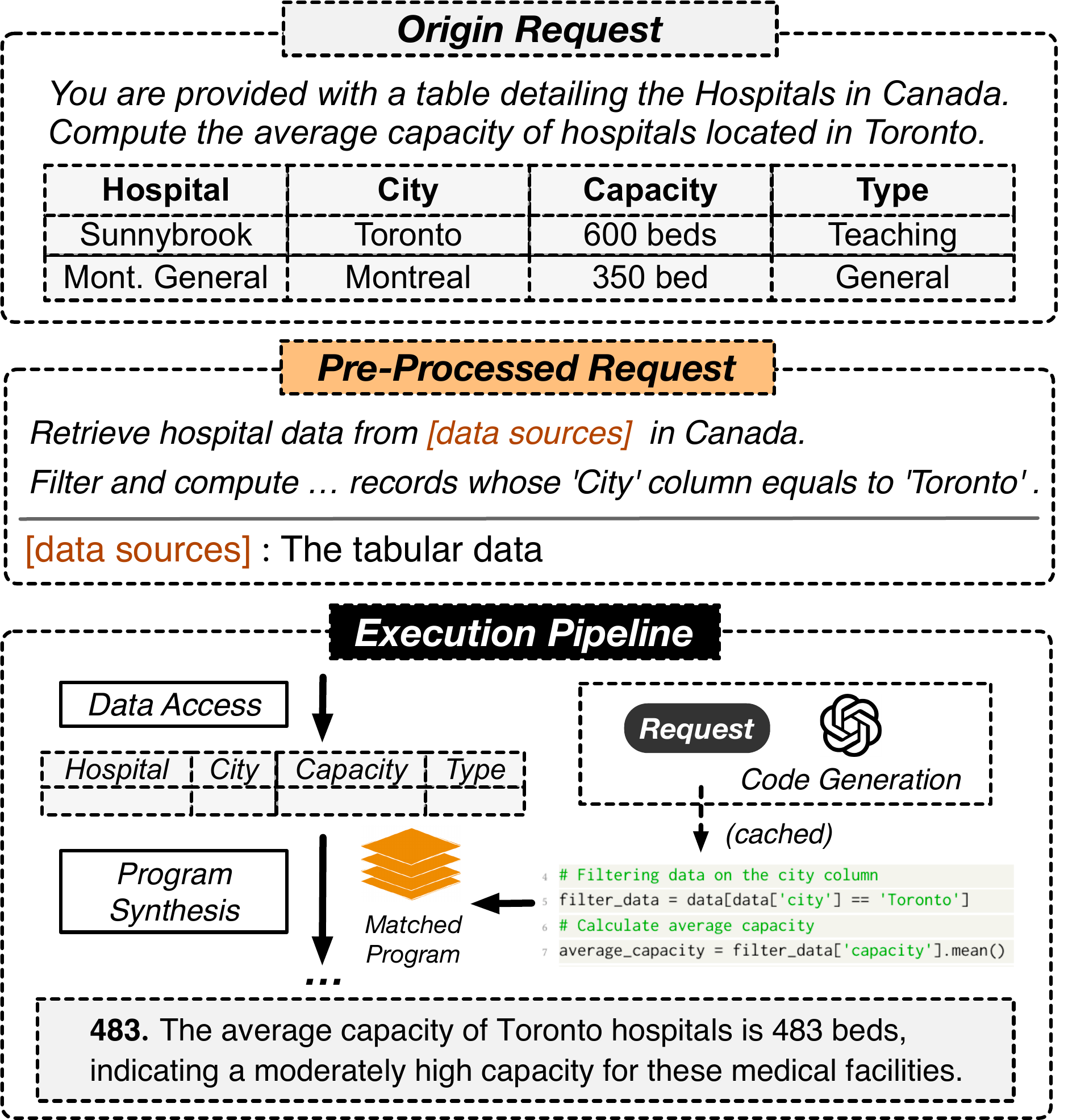}
	\vspace{-.5em}
	\caption{\oursys for NL-based Data Analytics}
	\label{fig:data-analytics}
	\vspace{-1.5em}
\end{figure}

\hi{Request Parsing.} The pipelines in data analytics are sequences of APIs tailored for the specific analytics task. Each chain in the pipeline represents a workflow, integrating multiple APIs to perform complex data analysis (e.g., data loading, transformation, and visualization). The operations of these chains utilizes the fine-tuned \llm to predict the most relevant API calls based on the input NL queries and data context (e.g., column correlations within the tabular data), ensuring accurate data processing flow. The following code demonstrates how to translate the natural language query into a data analytics workflow using Pandas. It involves loading the hospital data, filtering it by the condition ``located at Toronto'', and computing the average capacity:

\begin{lstlisting}[language=Python]
import pandas as pd
# Load data
data = pd.read_csv('hospital_data.csv')
# Filtering data on the city column
filter_data = data[data['city'] == 'Toronto']
# Calculate average capacity
average_capacity = filter_data['capacity'].mean()

print(f"The average capacity for hospitals located in Toronto is: {average_capacity}")
\end{lstlisting}

\hi{Execution Pipeline.} Apart from directly running the program derived by the above execution pipeline, we can utilize the vector database to function as a dynamic caching mechanism. That is, the vector database stores and manages the embeddings of natural language (NL) queries. These embeddings are effectively mapped to corresponding program or a sub-sequence of APIs that are essential for executing the analytics task at hand. For instance, when a query related to trend analysis within a given dataset is received, the vector database strategically caches embeddings. These cached embeddings are then mapped to a series of specialized APIs that are proficient in time-series analysis. This approach also enables the system to adaptively learn from incoming queries, thereby progressively optimizing the caching mechanism for future requests.

\hi{Research Challenges.} First, how to develop algorithms and models that can accurately generate analytical results from natural language queries. Second, how to automatically generate analysis programs, handling diverse data formats, and creating models capable of translating different programming languages to facilitate seamless data analytics. Third, how to efficiently optimize the sequence of API calls in data analytics workflows, i.e.,  developing algorithms to predict the most effective sequence of API calls.

\begin{figure}[!t]
	\vspace{.5em}
	\centering
	\includegraphics[width=.95\linewidth, trim={0 0 0 0},clip]{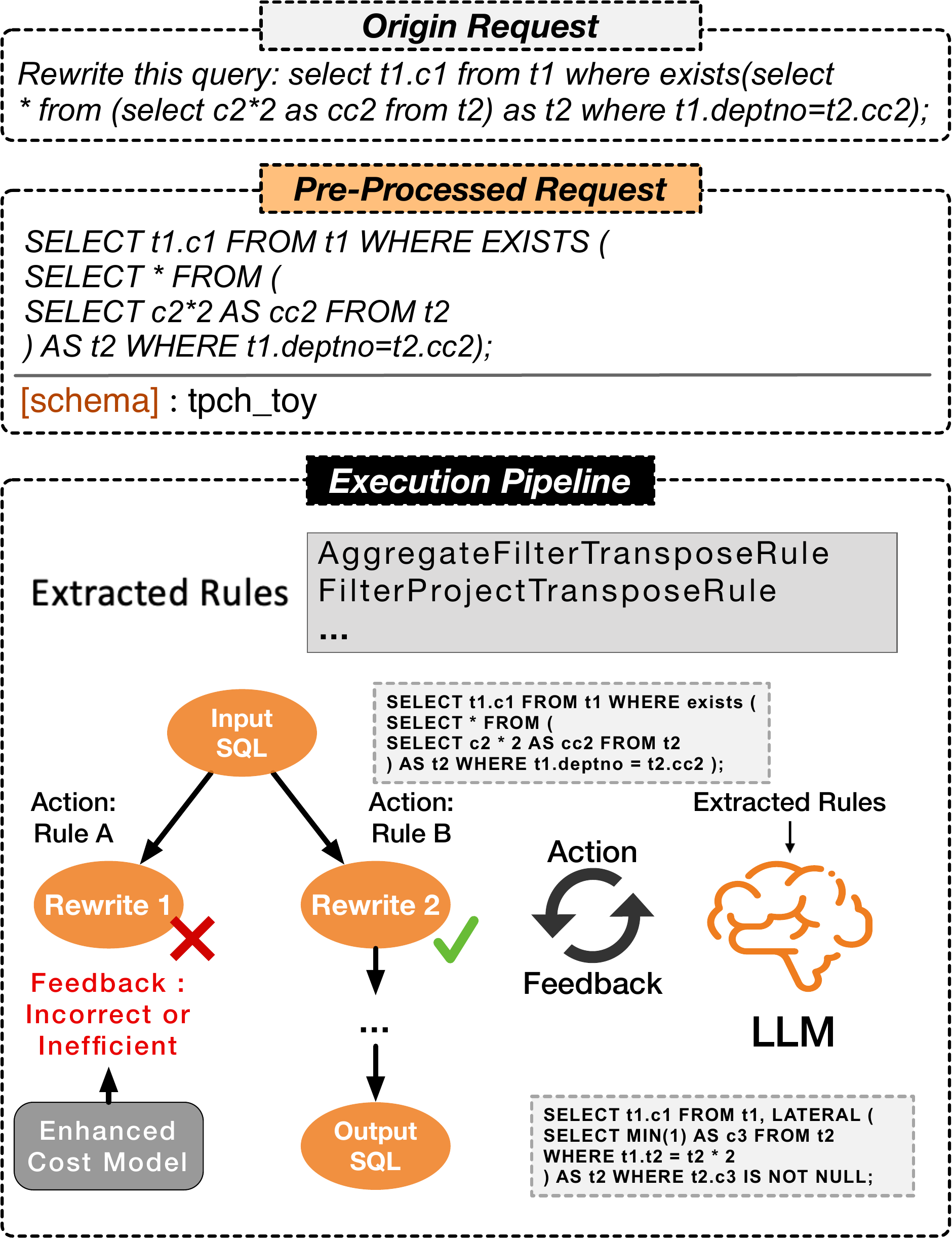}
	\vspace{-.75em}
	\caption{\oursys for Query Rewrite}
	\label{fig:queryrewrite}
	\vspace{-2.4em}
\end{figure}

\section{\oursys for Query Rewrite}
\label{sec:rewrite}

Query rewrite~\cite{volcano} aims to transfer a SQL query to an equivalent but more efficient SQL query. It includes two main types of work. The first type~\cite{DBLP:journals/pvldb/BaiA023,DBLP:conf/sigmod/WangZYDHDT0022}, aims to create new rewrite rules from existing SQL pairs~\cite{DBLP:conf/sigmod/WangZYDHDT0022} or to develop new domain-specific languages (DSLs) to make it easier to implement these rules~\cite{DBLP:journals/pvldb/BaiA023}. However, these methods are limited because they can only handle simple rules (e.g., with up to three operators); but they don't make use of valuable rewrite experiences in textual documents and DBA experiences. The second type of work~\cite{DBLP:journals/pvldb/ZhouLCF21,DBLP:journals/pvldb/Zhou0WLSZ23} tries to improve rewrite performance by changing the order of rule application. Yet, these methods also do not use structural information to $(i)$ reduce the search space or to $(ii)$ effectively judge the quality of rewrites at the logical query level. In this section, we showcase the methods and techniques used in \oursys for query rewrite.

\hi{Domain-Specific Data Preparation -- Rewrite Rules.} Given documents like database operation manuals or academic research papers, we can employ \llms to automatically \textit{extract, verify, and correct} optimization rules. These rules are formulated in a formal language, either as a logical DSL rule or as an executable function (e.g., Java). For each rule, we also create a natural language description. This description includes details like the rule's purpose, steps, application scenarios, and examples. These descriptions aid \llms in the executor agent to effectively rewrite queries.


\hi{Request Pre-Processing.} In most cases, extracting the target query from the rewrite requests is straightforward. This can be done by writing basic regular expression (regex) code to capture complete SQL statements. For example, we can use the regex pattern \textit{r``(select|$\cdots$|delete).*?;''} to retrieve queries ranging from ``select'' to ``delete'' commands.

\hi{Request Parsing.} We match the extracted SQL queries with the corresponding table schema and applicable rules. These rules may be well-formatted (executed by engines like Calcite~\cite{DBLP:conf/sigmod/BegoliCHML18} or 
CockroachDB~\cite{DBLP:conf/sigmod/TaftSMVLGNWBPBR20}), or require \llm in \emph{Execution Agent} to interpret and follow. Each rule is specifically designed to address different aspects of query rewrite. For example, one rule might aim at optimizing subqueries, while another might be focused on making aggregation processes more efficient. For each extracted query, we organize the sequence of matched rules as a basic execution pipeline. 

\hi{Executor Agent.} Within the execution pipeline, the \emph{Pipeline Refinement} module begins by selecting the most effective strategy of applying these rules. This involves training an order assignment \llm, which learns from experiences such as: $(i)$ rules involving different operators that can be applied simultaneously, and $(ii)$ removing aggregations within a subquery may enable the application of more effective rules that were previously inapplicable. Next, we evaluate the rewritten query by criteria like {\it whether it has effectively reduced the complexity of operators}. The final step is to adopt the rewritten query that ensures $(i)$ semantically equivalent results and $(ii)$ the lowest estimated cost.

\hi{Caching-based Execution.} We also support an advanced caching mechanism to enhance the rewrite procedure. After rewriting an SQL query, the vector database caches its embedding together with the rewrite strategies (i.e., rule sequence). This cached embedding is then leveraged to rapidly rewrite the following queries with similar rewrite preferences. For instance, if a cached embedding indicates a query with complex grouping, the vector database guides the system to employ the strategy optimized for grouping operations, thereby saving the rewrite time.

\hi{Model and Embedding Management.} During online usage, there can be  SQL queries where it has low confidence or makes incorrect predictions. For these queries, generate representative samples to $(i)$ train the \llm in \emph{Executor Agent} and $(ii)$ cache in vector databases, where each sample is composed of the original SQL query, rule, rewritten query, and explanatory annotations. Given the scarcity of high-quality samples for query rewriting~\cite{zhoudb}, we can apply methods like active learning to proactively collect such samples.

\hi{Research Challenges.} First, how to evaluate the quality and overlap of generated rules. Some rules may be redundant or represent similar conditions, which existing \llms may fail to identify. Second, how to utilize \llm to accurately verify semantic equivalence between the original and rewritten queries, which is a complex problem and lacks a general solution. Third, how to make a lightweight implementation within database kernels. This involves issues like minimizing the computational overhead of \llm inference and reducing the number of applied rules.

\vspace{-.5em}
\section{Conclusion}
\label{sec:conclusion}

In this paper, we propose an \llm-enhanced data management framework for addressing the limitations of \llms: hallucination, high cost, and low accuracy.  \oursys can well cover diversified data management applications. By integrating LLMs, vector database, and \llm agent, \oursys offers enhanced accuracy, relevance, and reliability of data processing. The framework’s potential in areas such as query rewriting, data analytics, and diagnosis is vast, promising significant advancements in various sectors. Part of the source code is open-sourced in {\it \href{https://github.com/TsinghuaDatabaseGroup/DB-GPT}{\textcolor{blue}{https://github.com/TsinghuaDatabaseGroup/DB-GPT}}} (and will soon be split off into an independent repository).




\begin{acks}
We thank Wei Zhou, Zhaoyan Sun, Zhiyuan Liu, and  \href{https://github.com/OpenBMB/AgentVerse}{\textcolor{black}{AgentVerse Team}} for their valuable advice on this vision. Zui Chen and Lei Cao have assisted in designing Figure~\ref{fig:queryrewrite}.
\end{acks}


\clearpage
\balance
\bibliographystyle{plain}
\bibliography{DA}

\begin{thebibliography}{10}

\bibitem{vitgpt}
https://huggingface.co/nlpconnect/vit-gpt2-image-captioning.
\newblock Last accessed on 2024-1.

\bibitem{Marcoroni}
https://huggingface.co/thebloke/marcoroni-7b-v3-gguf.
\newblock Last accessed on 2024-1.

\bibitem{DBLP:journals/pvldb/BaiA023}
Qiushi Bai, Sadeem Alsudais, and Chen Li.
\newblock Querybooster: Improving {SQL} performance using middleware services
  for human-centered query rewriting.
\newblock {\em Proc. {VLDB} Endow.}, 16(11):2911--2924, 2023.

\bibitem{DBLP:conf/sigmod/BegoliCHML18}
Edmon Begoli, Jes{\'{u}}s Camacho{-}Rodr{\'{\i}}guez, Julian Hyde, Michael~J.
  Mior, and Daniel Lemire.
\newblock Apache calcite: {A} foundational framework for optimized query
  processing over heterogeneous data sources.
\newblock In {\em {SIGMOD}}, pages 221--230. {ACM}, 2018.

\bibitem{DBLP:journals/tkde/ChaiLFL21}
Chengliang Chai, Guoliang Li, Ju~Fan, and Yuyu Luo.
\newblock Crowdchart: Crowdsourced data extraction from visualization charts.
\newblock {\em {IEEE} Trans. Knowl. Data Eng.}, 33(11):3537--3549, 2021.

\bibitem{chen2023seed}
Zui Chen, Lei Cao, Sam Madden, Tim Kraska, Zeyuan Shang, Ju~Fan, Nan Tang,
  Zihui Gu, Chunwei Liu, and Michael Cafarella.
\newblock Seed: Domain-specific data curation with large language models.
\newblock {\em arXiv e-prints}, pages arXiv--2310, 2023.

\bibitem{DBLP:conf/cidr/DiasRSVW05}
Karl Dias, Mark Ramacher, Uri Shaft, Venkateshwaran Venkataramani, and Graham
  Wood.
\newblock Automatic performance diagnosis and tuning in oracle.
\newblock In {\em Second Biennial Conference on Innovative Data Systems
  Research, {CIDR} 2005, Asilomar, CA, USA, January 4-7, 2005, Online
  Proceedings}, pages 84--94. www.cidrdb.org, 2005.

\bibitem{awesome-chatgpt-prompts}
Fatih Kadir~Akın et~al.
\newblock Awesome chatgpt prompts.
\newblock https://github.com/f/awesome-chatgpt-prompts, 2023.

\bibitem{enhance-chatgpt}
Jules~White et~al.
\newblock A prompt pattern catalog to enhance prompt engineering with chatgpt.
\newblock 2023.

\bibitem{prompt}
Pengfei~Liu et~al.
\newblock Pre-train, prompt, and predict: {A} systematic survey of prompting
  methods in natural language processing.
\newblock {\em {ACM} Comput. Surv.}, 2023.

\bibitem{gormley2015elasticsearch}
Clinton Gormley and Zachary Tong.
\newblock {\em Elasticsearch: the definitive guide: a distributed real-time
  search and analytics engine}.
\newblock " O'Reilly Media, Inc.", 2015.

\bibitem{volcano}
Goetz Graefe.
\newblock Volcano - an extensible and parallel query evaluation system.
\newblock {\em {IEEE} Trans. Knowl. Data Eng.}, 6(1):120--135, 1994.

\bibitem{hong2023metagpt}
Sirui Hong, Xiawu Zheng, Jonathan Chen, Yuheng Cheng, Ceyao Zhang, Zili Wang,
  Steven Ka~Shing Yau, Zijuan Lin, Liyang Zhou, Chenyu Ran, et~al.
\newblock Metagpt: Meta programming for multi-agent collaborative framework.
\newblock {\em arXiv preprint arXiv:2308.00352}, 2023.

\bibitem{DBLP:journals/pvldb/MaYZWZJHLLQLCP20}
Minghua Ma, Zheng Yin, Shenglin Zhang, and et~al.
\newblock Diagnosing root causes of intermittent slow queries in large-scale
  cloud databases.
\newblock {\em Proc. {VLDB} Endow.}, 13(8):1176--1189, 2020.

\bibitem{marquez2008semantic}
Llu{\'\i}s M{\`a}rquez, Xavier Carreras, Kenneth~C Litkowski, and Suzanne
  Stevenson.
\newblock Semantic role labeling: an introduction to the special issue, 2008.

\bibitem{qian2023communicative}
Chen Qian, Xin Cong, Cheng Yang, Weize Chen, Yusheng Su, and et~al.
\newblock Communicative agents for software development.
\newblock {\em arXiv preprint arXiv:2307.07924}, 2023.

\bibitem{qin2023tool}
Yujia Qin, Shengding Hu, Yankai Lin, and et~al.
\newblock Tool learning with foundation models.
\newblock {\em arXiv preprint arXiv:2304.08354}, 2023.

\bibitem{qin2023toolllm}
Yujia Qin, Shihao Liang, Yining Ye, Kunlun Zhu, Lan Yan, Yaxi Lu, Yankai Lin,
  Xin Cong, Xiangru Tang, Bill Qian, Sihan Zhao, Runchu Tian, Ruobing Xie, Jie
  Zhou, Mark Gerstein, Dahai Li, Zhiyuan Liu, and Maosong Sun.
\newblock Toolllm: Facilitating large language models to master 16000+
  real-world apis, 2023.

\bibitem{toolformer}
Timo Schick, Jane Dwivedi{-}Yu, Roberto Dess{\`{\i}}, Roberta Raileanu, Maria
  Lomeli, Luke Zettlemoyer, Nicola Cancedda, and Thomas Scialom.
\newblock Toolformer: Language models can teach themselves to use tools, 2023.

\bibitem{reflexion}
Noah Shinn, Beck Labash, and Ashwin Gopinath.
\newblock Reflexion: an autonomous agent with dynamic memory and
  self-reflection, 2023.

\bibitem{DBLP:conf/sigmod/TaftSMVLGNWBPBR20}
Rebecca Taft, Irfan Sharif, Andrei Matei, Nathan VanBenschoten, and et~al.
\newblock Cockroachdb: The resilient geo-distributed {SQL} database.
\newblock In {\em {SIGMOD}}, pages 1493--1509. {ACM}, 2020.

\bibitem{tang2020rpt}
Nan Tang, Ju~Fan, Fangyi Li, Jianhong Tu, Xiaoyong Du, Guoliang Li, Sam Madden,
  and Mourad Ouzzani.
\newblock Rpt: relational pre-trained transformer is almost all you need
  towards democratizing data preparation.
\newblock {\em arXiv preprint arXiv:2012.02469}, 2020.

\bibitem{DBLP:conf/coling/Turney08}
Peter~D. Turney.
\newblock A uniform approach to analogies, synonyms, antonyms, and
  associations.
\newblock In Donia Scott and Hans Uszkoreit, editors, {\em {COLING}}, pages
  905--912, 2008.

\bibitem{DBLP:conf/sigmod/WangZYDHDT0022}
Zhaoguo Wang, Zhou Zhou, Yicun Yang, Haoran Ding, Gansen Hu, Ding Ding, Chuzhe
  Tang, Haibo Chen, and Jinyang Li.
\newblock Wetune: Automatic discovery and verification of query rewrite rules.
\newblock In {\em {SIGMOD}}, pages 94--107. {ACM}, 2022.

\bibitem{cot}
Jason Wei, Xuezhi Wang, Dale Schuurmans, Maarten Bosma, Brian Ichter, Fei Xia,
  Ed~Chi, Quoc Le, and Denny Zhou.
\newblock Chain-of-thought prompting elicits reasoning in large language
  models, 2023.

\bibitem{wei2022chain}
Jason Wei, Xuezhi Wang, Dale Schuurmans, Maarten Bosma, Fei Xia, Ed~Chi, Quoc~V
  Le, Denny Zhou, et~al.
\newblock Chain-of-thought prompting elicits reasoning in large language
  models.
\newblock {\em Advances in Neural Information Processing Systems},
  35:24824--24837, 2022.

\bibitem{react}
Shunyu Yao, Jeffrey Zhao, Dian Yu, Nan Du, Izhak Shafran, Karthik Narasimhan,
  and Yuan Cao.
\newblock React: Synergizing reasoning and acting in language models, 2023.

\bibitem{DBLP:conf/sigmod/YoonNM16}
Dong~Young Yoon, Ning Niu, and Barzan Mozafari.
\newblock Dbsherlock: {A} performance diagnostic tool for transactional
  databases.
\newblock In Fatma {\"{O}}zcan, Georgia Koutrika, and Sam Madden, editors, {\em
  Proceedings of the 2016 International Conference on Management of Data,
  {SIGMOD} Conference 2016, San Francisco, CA, USA, June 26 - July 01, 2016},
  pages 1599--1614. {ACM}, 2016.

\bibitem{DBLP:journals/pvldb/ZhouLCF21}
Xuanhe Zhou, Guoliang Li, Chengliang Chai, and Jianhua Feng.
\newblock A learned query rewrite system using monte carlo tree search.
\newblock {\em Proc. {VLDB} Endow.}, 15(1):46--58, 2021.

\bibitem{zhou2023llm}
Xuanhe Zhou, Guoliang Li, and Zhiyuan Liu.
\newblock Llm as dba.
\newblock {\em arXiv preprint arXiv:2308.05481}, 2023.

\bibitem{DBLP:journals/corr/abs-2312-01454}
Xuanhe Zhou, Guoliang Li, Zhaoyan Sun, Zhiyuan Liu, Weize Chen, Jianming Wu,
  Jiesi Liu, Ruohang Feng, and Guoyang Zeng.
\newblock D-bot: Database diagnosis system using large language models.
\newblock {\em CoRR}, abs/2312.01454, 2023.

\bibitem{DBLP:journals/pvldb/Zhou0WLSZ23}
Xuanhe Zhou, Guoliang Li, Jianming Wu, Jiesi Liu, Zhaoyan Sun, and Xinning
  Zhang.
\newblock A learned query rewrite system.
\newblock {\em Proc. {VLDB} Endow.}, 16(12):4110--4113, 2023.

\bibitem{zhoudb}
Xuanhe Zhou, Zhaoyan Sun, and Guoliang Li.
\newblock Db-gpt: Large language model meets database.

\end{thebibliography}
\balance



\end{document}